\documentclass[conference,a4paper]{IEEEtran}
\IEEEoverridecommandlockouts %this enabbles thanks in the conference mode
\usepackage[T1]{fontenc}
\usepackage{amsmath}
\usepackage{bm}

\usepackage{amsmath,amssymb}
\usepackage{graphicx}
\usepackage{tikz}
\usetikzlibrary{positioning} % below=of
\usepackage{microtype}
\usepackage[space]{cite}
\usepackage{subcaption}
\newcommand{\tail}[1]{{\small\textsf{tail}}(#1)}
\newcommand{\head}[1]{{\small\textsf{head}}(#1)}
\newcommand{\outset}[1]{{\small\textsf{out}}({#1})}
\newcommand{\inset}[1]{{\small\textsf{in}}({#1})}

\newtheorem{theorem}{Theorem}

\newtheorem{remark}{Remark}

\graphicspath{{.}{../pictures/}}

\def\gap{1.01ex}
\abovedisplayskip\gap
\belowdisplayskip\gap
\abovedisplayshortskip\gap
\belowdisplayshortskip\gap

\begin{document}

%---------- Title ----------
\title{An Equivalence Between\\ Secure Network and Index Coding}
\author{\IEEEauthorblockN{Lawrence Ong$^\dagger$, Badri N.\ Vellambi$^\ddagger$, J\"{o}rg Kliewer$^\ddagger$, and Phee Lep Yeoh$^{\S}$}
\IEEEauthorblockA{$^\dagger$The University of Newcastle, Australia; $^\ddagger$New Jersey Institute of Technology, USA; $^\S$University of Sydney, Australia}
\thanks{This work is supported by ARC grants FT140100219, DE140100420, and DP150100903, and US NSF grants CNS-1526547 and CCF-1439465.}
%\vspace{-1.7ex}
}
%\vspace*{-3ex}
\maketitle
%\vspace*{-4ex}

\begin{abstract}
We extend the equivalence between network coding and index coding by Effros, El Rouayheb, and Langberg to the secure communication setting in the presence of an eavesdropper. Specifically, we show that the most general versions of secure network-coding setup by Chan and Grant and the secure index-coding setup by Dau, Skachek, and Chee, which also include the randomised encoding setting,  are equivalent.
\end{abstract}

%%%%
%%%%
%%%%
%%%%
%\vspace{-3ex}
\section{Introduction}

Recently, equivalence results in information theory and network coding have been of significant interest in the community. Such reduction
results uniquely map one communication problem  to another equivalent problem  that is potentially easier to study than the original
problem. Some of the equivalence results already established include those between instances of multiple-unicast network coding and those of (a) multiple-multicast network coding~\cite{DoughertyZeger06}, (b) secure network coding~\cite{Huangetal13}, and (c) index coding~\cite{effrosrouayheblangberg15}.

In particular, the latter result addresses the equivalence between network
coding~\cite{ahlswedecai00} and index coding~\cite{baryossefbirk11} in  the non-secure setting. Non-secure network
coding and index coding were
shown to be
equivalent~\cite{rouayhebsprintsongeorghiades10,effrosrouayheblangberg15} in
the sense that a network-coding instance can be mapped to an equivalent
index-coding instance, for which a code for one instance can be translated
to the other, and vice versa. Similarly, an index-coding instance can be
mapped to an equivalent network-coding instance, with a suitable code
translation. 

While strongly-secure and weakly-secure network coding~\cite{caiyeung11,bhattadnarayanan05} as well as strongly-secure and
weakly-secure index coding~\cite{dauskachekchee12} have been studied in the
literature so far, an equivalence between these coding approaches for the
\emph{secure} setting has not been addressed  to the best
of our knowledge.

Note that the  equivalence between non-secure network and index coding does
not trivially apply to the secure setting. In particular, we pointed out \cite{ongvellambiyeohklieweryuan2016} that equating the eavesdropper settings in secure network coding and secure index coding is not straightforward. We also showed that the equivalence breaks down in the randomised encoding setting---noting that randomised encoding is inevitable in some secure network-coding instances~\cite{caiyeung11}.

% In the literature, strongly-secure and weakly-secure network coding have been studied~\cite{caiyeung11,bhattadnarayanan05}, as well as strongly-secure and weakly-secure index coding~\cite{dauskachekchee12}. However, there is no known equivalence between secure network and index coding.

% In the classical non-secure setting, network coding~\cite{ahlswedecai00} and index coding~\cite{baryossefbirk11} were shown to be equivalent~\cite{rouayhebsprintsongeorghiades10,effrosrouayheblangberg15} in the sense that a network-coding instance can be mapped to an equivalent index-coding instance, for which a code for one instance can be translated to the other, and vice versa. Similarly, an index-coding instance can be mapped to an equivalent network-coding instance, with a suitable code translation.

% The classical equivalence between network and index coding does not trivially apply to the secure setting. In particular, we pointed out~\cite{ongvellambiyeohklieweryuan2016} that equating the eavesdropper settings in secure network coding and secure index coding is not straightforward. We also showed that the equivalence breaks down in the randomised encoding setting---noting that randomised encoding is inevitable in some secure network-coding instances~\cite{caiyeung11}.

In this paper, we extend the equivalence between network and index coding by Effros, El Rouayheb, and Langberg~\cite{effrosrouayheblangberg15} to strongly-secure and weakly-secure settings, by proposing a suitable mapping for the eavesdroppers. For the mapping from secure network coding to secure index coding, we also introduce the concept of an augmented secure network-coding instance to capture the randomness in the encoding. With this, we incidentally establish an equivalence between non-secure network and index coding with randomised encoding.

%\vspace{-1ex}

\section{Problem Definition and Notation}

For any positive integer $n$, let $[n]\triangleq \{1,\ldots,n\}$. For a set $\mathcal{I} = \{i_1, \dotsc, i_{|\mathcal{I}|}\}$, let $\boldsymbol{X}_\mathcal{I} \triangleq [X_{i_1} \dotsm X_{i_{|\mathcal{I}|}}]$ with an arbitrary but fixed order. The tail and head of an edge $(u,v)\in E$ in a directed graph $G=(V,E)$ refer to vertices $u$, and $v$, respectively, i.e., $u=\tail{e}$ and $v=\head{e}$. For a node $v\in V$, we let $\inset{v}$ to be the set of all edges with $\head{e}=v$; similarly, $\outset{v}$ denotes the set of all edges with $\tail{e}=v$.

\subsection{Secure network coding}

\subsubsection{Network-coding instances}
We follow Chan and Grant's secure network-coding definition~\cite{changrant08}. It includes Bhattad and Narayanan's weakly-secure network-coding definition~\cite{bhattadnarayanan05} and Cai and Yeung's strongly-secure network coding definition~\cite{caiyeung11} as special cases.
A secure network-coding instance, denoted by $\mathcal{I} = ( G, M, W)$, is defined as follows:
\begin{itemize}
\item $G = (\mathcal{V}, \mathcal{E})$ is an acyclic graph with vertex set $\mathcal{V}$ and edge set $\mathcal{E}$. Each edge $e \in \mathcal{E}$ has a \textit{capacity} given by $c_e$.
\item $M = (\mathcal{S}, O, \mathcal{D})$ is the connection requirement. The set $\mathcal{S}$ is the collection of source-message indices, where the source messages $\{X_s: s \in \mathcal{S}\}$ are mutually independent and are each distributed on $[2^{R_s n}]$, for some positive integer $n$ that can be chosen to suit the design of network codes. Here, $R_s$ denotes the rate of the message $X_s$, $s\in \mathcal{S}$. The source-location mapping $O: \mathcal{S} \rightarrow \mathcal{V}$ specifies the originating node $O(s)$ for the source message $X_s$. The destination-location mapping $\mathcal{D}: \mathcal{S} \rightarrow 2^{\mathcal{V}}$ specifies the nodes $\mathcal{D}(s)$ that require the message $X_s$.
  \item $W = ((\mathcal{A}_r,\mathcal{B}_r):r \in \mathcal{R})$ defines the eavesdropping pattern for $|\mathcal{R}|$ eavesdroppers. Each eavesdropper $r \in \mathcal{R}$ observes the set of links $\mathcal{B}_r \subseteq \mathcal{E}$ and tries to reconstruct a subset of source messages indexed by $\mathcal{A}_r \subseteq \mathcal{S}$, i.e., $\bm{X}_{\mathcal{A}_r}$.
  \end{itemize}

 We assume that vertices with no incoming links are originating nodes for some source messages, and  vertices with no outgoing links are destinations for some source messages.

 \subsubsection{Deterministic network codes}
 Given $(G,M)$, a network code $(\mathcal{F},\mathcal{G})$ consists of a collection of encoding functions for the edges $\mathcal{F} = \{f_e:e \in \mathcal{E}\}$, and decoding functions for the vertices $\mathcal{G} = \{g_u: u \in \mathcal{V}\}$ satisfying the following:

The local encoding function $f_e$ for edge $e$  takes in random variables associated with $\inset{\tail{e}}$ and source messages originating at node $\tail{e}$, and  outputs a random variable associated with link~$e$, denoted by $X_e \in [2^{c_e n}]$. %$\{f_e\}$ are commonly called the local encoding function.

Given that $G$ is acyclic, each edge message $X_e$ can be written as a function of source messages originating from its predecessors, denoted by $\bar{f}_e$. This is known as the global encoding function, and it can be recursively calculated (following the topology of the graph) using (i) $\bar{f}_e = f_e$ if $\tail{e}$ has no incoming links, and (ii) $\bar{f}_e = f_e(\bar{f}_{e_1}, \bar{f}_{e_2}, \dotsc, \bar{f}_{e_n})$, where $\{e_1, e_2, \dotsc, e_n\} = \inset{\tail{e}}$.

The decoding function $g_u$ for a node~$u \in \mathcal{V}$ takes in random variables associated with links $\inset{u}$ and source messages originating at node $u$, and outputs  $\bm{X}_{\{s \in \mathcal{S}: u \in \mathcal{D}(s)\}}$.
In this paper, we only consider \textit{zero-error} decoding.

\subsubsection{Randomised network codes}
A network code is said to be randomised if there exists an edge function $f_e$ that is not a deterministic function of the random variables associated with $\inset{\tail{e}}$ and source messages originating at node $\tail{e}$.

Any randomised function can be implemented by generating an independent random variable $Z_u$ at each node $u \in \mathcal{V}$, where $Z_u$ takes values in an alphabet with size $\prod_{e \in \outset{u}} 2^{c_e n}$, where $\outset{u}$ is defined as the set of all outgoing edges from node $u$.
These independent random variables are often referred to as \textit{random keys}.

A randomised network code $(\mathcal{F}',\mathcal{G})$ is similar to a deterministic network code $(\mathcal{F},\mathcal{G})$, except that each edge encoding function $f_e'$ is a function of (i) random variables associated with $\inset{\tail{e}}$, (ii) source messages originating at node $\tail{e}$, and (iii) the random key $Z_{\tail{e}}$.

\subsubsection{Secure network codes}
A deterministic or randomised network code $(\mathcal{F},\mathcal{G})$ for $(G,M)$ is said to be secure against an eavesdropping pattern $W$ if each eavesdropper $r$ gains no information about  $\bm{X}_{\mathcal{A}_r}$ that it attempts to reconstruct after observing $\bm{X}_{\mathcal{B}_r}$ on the links it has access to, i.e.,
\begin{equation}
  H(\bm{X}_{\mathcal{A}_r}| \bm{X}_{\mathcal{B}_r}) = H(\bm{X}_{\mathcal{A}_r}), \quad r\in \mathcal{R}. \label{eq:secure-nc-condition}
\end{equation}
In other words, $(\mathcal{F},\mathcal{G})$ is a secure network code for the secure network-coding instance $I$.

\subsubsection{Secure network-coding rates}
The secure network-coding instance is said to be $(\bm{R}_{\mathcal{S}},n)$-feasible if and only if there exists at least one secure network code with the associated source-message rates and block size $n$.

% \begin{remark}
%   We have used the network-coding definition by Chan and Grant~\cite{changrant08} instead of the definition by Effros, El Rouayheb, and Langberg~\cite{effrosrouayheblangberg15}, although they are equivalent as far as classical network coding is concerned (i.e., without security constraints). In the definition by Effros et al., each source message is generated by one \textit{source node} without any incoming links. The definition by Chan and Grant, used in this paper, can be easily converted to that by Effros et al.\ as follows: For each node~$u$ that is (i) generating multiple source messages or (ii) generating source message(s) and having incoming link(s), we remove the source messages at $u$ and add source nodes each generating a source message and each with a link to $u$. However, by doing so, we change the min-cut of the message flows in the network, thereby making the secure-network-coding results incompatible with that in Cai and Yeung~\cite{caiyeung11}.
% \end{remark}

\subsection{Secure index coding}

\subsubsection{Secure index-coding instances}
We follow Dau, Skachek, and Chee's secure index-coding definition~\cite{dauskachekchee12}. A secure index-coding instance, denoted by $\hat{I} = (\hat {\mathcal{S}}, \hat{\mathcal{T}}, \{\hat{\mathcal{W}}_{\hat{t}}\}, \{ \hat{\mathcal{H}}_{\hat{t}} \}, \hat{W})$, is defined as follows:
\begin{itemize}
\item $\hat{\mathcal{S}}=[k]$ is the set of indices of $k$ source messages available at a sender. The messages $\{\hat{X}_{\hat{s}}: \hat{s} \in \hat{\mathcal{S}}\}$ are mutually independent and for $\hat s \in \hat{ \mathcal{S}}$,  $\hat X_{\hat s}$ is distributed on $[2^{\hat{R}_{\hat{s}}n}]$, for some non-negative message rate $\hat R_{\hat s}$ and positive integer $n$ that can be chosen to suit the design of index codes.
\item $\hat{\mathcal{T}} =[\ell]$ is the collection of $\ell$ receiver indices.
\item $\hat{\mathcal{W}}_{\hat{t}}$ is the set of the indices of the messages required by receiver~$\hat{t} \in \hat{\mathcal{T}}$.
\item $\hat{\mathcal{H}}_{\hat{t}}$ is the set of indices of the messages known a priori to receiver~$\hat{t} \in \hat{\mathcal{T}}$.
  \item $\hat{W} = ((\hat{\mathcal{A}}_{\hat{r}},\hat{\mathcal{B}}_{\hat{r}}):\hat{r} \in \hat{\mathcal{R}})$ is the eavesdropping pattern. % \footnote{The notation used by Dau, Skachek, and Chee is reversed, i.e., $\mathcal{A}$ for the indices of messages the eavesdropper can access and $\mathcal{B}$ for the indices of messages the eavesdropper tries to reconstruct. Here, we have used the notation that mirrors that for secure network coding.}
    Each eavesdropper~$\hat{r}\in \hat{\mathcal{R}}$ has access to the codeword broadcast by the sender and a subset of the messages $\bm{X}_{\hat{\mathcal{B}}_{\hat{r}}}$, and tries to reconstruct $\bm{X}_{\hat{\mathcal{A}}_{\hat{r}}}$, where $\hat{\mathcal{A}}_{\hat{r}},\hat{\mathcal{B}}_{\hat{r}} \subseteq \hat{\mathcal{S}}$.
\end{itemize}

\subsubsection{Deterministic index codes}
A deterministic index code $(\hat{\mathcal{F}},\hat{\mathcal{G}}) =
(\hat{f},\{\hat{g}_{\hat{t}}:\hat{t} \in \hat{\mathcal{T}} \})$ consists of
an encoding function by the sender $\hat{\mathcal{F}}=\hat{f}$ which takes
in the random variables $\hat{\bm{X}}_{\hat{\mathcal{S}}}$ and outputs a
random variable $\hat{X}_\text{b} \in [2^{\hat{c}_\text{b}n}]$, where $\hat{c}_\text{b}$ is the broadcast rate, and $n$ is the block size of the code. It also
consists of a decoding function $\hat{g}_{\hat{t}}$ for receiver $\hat{t}$,
which takes in the sender's codeword $\hat{X}_\text{b}$ and its prior
messages $\hat{\bm{X}}_{\hat{\mathcal{H}}_{\hat{t}}}$ and outputs  the
messages $\hat{\bm{X}}_{\hat{\mathcal{W}}_{\hat{t}}}$ it requires.
Similar to the network-coding setup, we only consider zero-error decoding.

\subsubsection{Randomised index codes}
A randomised index code $(\hat{\mathcal{F}}',\hat{\mathcal{G}})$ is defined similar to the deterministic index codes except that the sender's encoding function takes in an independent random key $\hat{Z} \in [2^{\hat{r}_\text{b}n}]$ of some positive rate $\hat{r}_\text{b}$ in addition to $\hat{\bm{X}}_{\hat{\mathcal{S}}}$. 
Unlike the model by Mojahedian, Aref, and Gohari~\cite{mojahediangohariaref15}, the randomness allowed in the encoding in our setting  is known only to the sender, and is not shared between the sender and the receivers.

\subsubsection{Secure index codes}
A deterministic or randomised index code
$(\hat{\mathcal{F}},\hat{\mathcal{G}})$ is said to be secure against the
eavesdropping pattern $\hat{W}$ if no eavesdropper $\hat{r}$ gains no information
about the message set $\hat{\bm{X}}_{\hat{\mathcal{A}}_{\hat{r}}}$ it tries to reconstruct by observing the
sender's codeword $\hat{X}_{\text{b}}$ and its side information $\hat{\bm{X}}_{\hat{\mathcal{B}}_{\hat{r}}}$, i.e.,
\begin{equation}
  H(\hat{\bm{X}}_{\hat{\mathcal{A}}_{\hat{r}}} | \hat{X}_{\text{b}}, \hat{\bm{X}}_{\hat{\mathcal{B}}_{\hat{r}}}) = H(\hat{\bm{X}}_{\hat{\mathcal{A}}_{\hat{r}}}),\quad \hat{r} \in \hat{\mathcal{R}}.
\end{equation}
 Clearly, $\hat{\mathcal{A}}_{\hat{r}} \cap \hat{\mathcal{B}}_{\hat{r}} = \emptyset$.
Specifically, we say that $(\hat{\mathcal{F}},\hat{\mathcal{G}})$ is  a secure index code for the secure index-coding instance $\hat{I}$.

\subsubsection{Secure index-coding rate}
The secure index-coding instance is said to be $(\hat{\bm{R}}_{\hat{\mathcal{S}}},\hat{c}_{\text{b}},n)$-feasible if and only if there exists at least one secure index code with the associated source-message rates $\hat{\bm{R}}_{\hat{\mathcal{S}}}$, broadcast rate $\hat{c}_{\text{b}}$, and block size $n$.

\section{Mapping from Secure Index Coding to\\ Secure Network Coding}

\begin{figure*}
  \centering
  \begin{subfigure}[b]{0.45\textwidth}
    \centering
        \resizebox{40ex}{!}{%
\includegraphics{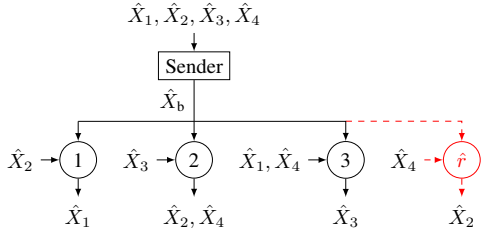}
}
        \caption{A secure index-coding instance $\hat{I}$, where an eavesdropper~$\hat{r}$ has access to the broadcast message $\hat{X}_{\text{b}}$, side information $\hat{X}_4$, and tries to reconstruct $\hat{X}_2$}
        \label{fig:index-coding-01}
      \end{subfigure}
      \quad
    \begin{subfigure}[b]{0.45\textwidth}
      \centering
      \resizebox{30ex}{!}{%
\includegraphics{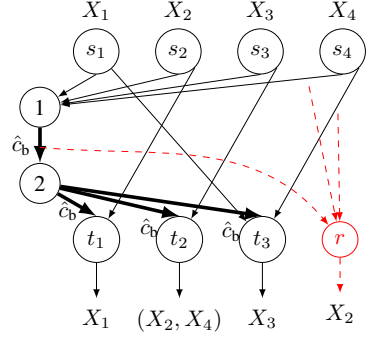}
}
        \caption{A secure network-coding instance $I$, where an eavesdropper~$r$ has access to link $(1,2)$,  all outgoing links from node $s_4$, and tries to reconstruct $X_2$. The capacity of all links given by thick arrows is $\hat{c}_{\text{b}}$}
        \label{fig:network-coding-01}
      \end{subfigure}
      \caption{A secure index-coding instance $\hat{I}$ and its
        corresponding secure network-coding instance $I$}
      \vspace{-2ex}
      \label{fig:index-to-network-example}
\end{figure*}

Given an instance $\hat{I} = (\hat {\mathcal{S}}, \hat{\mathcal{T}}, \{\hat{\mathcal{W}}_{\hat{t}}\}, \{ \hat{\mathcal{H}}_{\hat{t}} \}, \hat{W} )$ of a secure index-coding problem,  where $\hat{\mathcal{S}}= [k]$ and $\hat{\mathcal{T}} = [\ell]$, we construct an equivalent secure network-coding instance $\mathcal{I} = ( G, M, W)$ using the following rule:\\
\noindent \underline{Index-to-network coding mapping}:
\begin{itemize}
\item The graph $G = (\mathcal{V},\mathcal{E})$ consists of $k+\ell+2$ vertices labelled as $\mathcal{V}=\{s_1, s_2, \dotsc, s_{k}, t_1, t_2, \dotsc, t_{\ell}, 1, 2\}$. For each $i \in \hat{\mathcal{S}}$, vertex $s_i$ has an outgoing link to vertex~1 and to each vertex in $\{t_j: i \in \hat{\mathcal{H}_j} \}$. Each of these links from vertex $s_i$ are of capacity $2^{\hat{R}_i n}$. Vertex~1 has a link of capacity $2^{\hat{c}_{\text{b}}n}$ to vertex~2 and to each vertex in $\{t_i: i \in \hat{\mathcal{T}}\}$.
\item The connection requirement $M$ consists of the following: $\mathcal{S} = \hat{\mathcal{S}}$, and $\bm{X}_\mathcal{S}$ has the same distribution as $\hat{\bm{X}}_{\hat{\mathcal{S}}}$, which implies $R_i = \hat{R}_i$.
For each message $X_i$, $i \in \mathcal{S}$, the source locations are $O(i) = s_i$, i.e., the message $X_i$ originates at vertex~$s_i$, and is destined for $\mathcal{D}(i) = \{t_j: i \in \hat{\mathcal{W}}_j \}$.
  \item The eavesdropping pattern $W$ is defined as $\mathcal{R} = \hat{\mathcal{R}}$, $\mathcal{B}_r = \{ (1 \rightarrow 2), \{\outset{s_i} : i \in \hat{\mathcal{B}}_r \}\}$, and $\mathcal{A}_r = \hat{\mathcal{A}}_r$.
  \end{itemize}

  Note that by construction, for each $i \in \hat{\mathcal{T}}$,
  \begin{itemize}
  \item $\hat{\mathcal{W}}_i = \{j \in \mathcal{S}: t_i \in \mathcal{D}(j) \}$.
  \item $\hat{\mathcal{H}}_i = \{ j \in \mathcal{S}: (s_j \rightarrow t_i) \in \mathcal{E}\}$.
    \item The vertices in $\mathcal{V} \setminus \{t_1, \dotsc, t_{\ell} \}$ are not the destinations of any source message.
    \end{itemize}

Figure~\ref{fig:index-to-network-example} depicts an example of such a mapping. With the above conversion, we now state an equivalence between these two instances:
  \begin{theorem}\label{theorem:index-to-network}
    Let $\hat{I}$ and $\hat{c}_{\text{b}}$ be a secure index-coding instance and a broadcast rate, respectively. Let $I$ be the corresponding secure network-coding instance using the index-to-network coding mapping. For any $\hat{\bm{R}}_{\hat{\mathcal{S}}}$ and $\hat{c}_{\text{b}}$, the instance $\hat{I}$ is $(\hat{\bm{R}}_{\hat{\mathcal{S}}},\hat{c}_{\text{b}},n)$-feasible if and only if $I$ is $(\bm{R}_{\mathcal{S}},n)$-feasible.
  \end{theorem}

  \begin{IEEEproof}[Proof of Theorem~\ref{theorem:index-to-network}]

    \noindent \underline{$\hat{I}$ is $(\hat{\bm{R}}_{\hat{\mathcal{S}}},\hat{c}_{\text{b}},n)$-feasible $\Rightarrow$ $I$ is $(\bm{R}_{\mathcal{S}},n)$-feasible:}

    Let $(\hat{\mathcal{F}},\hat{\mathcal{G}})$ be a secure index code that supports $(\hat{\bm{R}}_{\hat{\mathcal{S}}},\hat{c}_{\text{b}},n)$. Then,    \begin{align}
      &H(\hat{\bm{X}}_{\hat{\mathcal{W}}_{\hat{t}}} | \hat{X}_{\text{b}}, \hat{\bm{X}}_{\hat{\mathcal{H}}_{\hat{t}}}) = 0, \quad &\text{for all } \hat{t} \in \hat{\mathcal{T}}, \label{eq:dedocability} \\
  &H(\hat{\bm{X}}_{\hat{\mathcal{A}}_{\hat{r}}} | \hat{X}_{\text{b}}, \hat{\bm{X}}_{\hat{\mathcal{B}}_{\hat{r}}}) = H(\hat{\bm{X}}_{\hat{\mathcal{A}}_{\hat{r}}}), &\text{for all } \hat{r} \in \hat{\mathcal{R}}, \label{eq:security}
    \end{align}
    where $\hat{X}_{\text{b}} = \hat{f}(\hat{\bm{X}}_{\hat{\mathcal{S}}}, \hat{Z})$, where $\hat{Z} \in [2^{\hat{c}_{\text{b}}n}]$ is a random key independent of the source messages (to account for randomised index coding), and
    \begin{equation}
      \hat{\bm{X}}_{\hat{\mathcal{W}}_{\hat{t}}} = \hat{g}_{\hat{t}}(\hat{f}(\hat{\bm{X}}_{\hat{\mathcal{S}}},\hat{Z}),\hat{\bm{X}}_{\hat{\mathcal{H}}_{\hat{t}}}), \quad\quad\quad \text{for all } \hat{t} \in \hat{\mathcal{T}}. \label{eq:decoding}
    \end{equation}
%    for all $\hat{t} \in \hat{\mathcal{T}}$.

    Now, we construct a secure network code as follows:
    \begin{itemize}
    \item Set $f_e = X_{\tail{e}}$ for each outgoing edge $e$ from each vertex in $\{s_i: i \in \hat{\mathcal{S}}\}$. This is possible since vertex~$s_i$ is the originating vertex for the message~$X_i$.
    \item Set $f_{(1, 2)} = f_e = \hat{f}(\bm{X}_{\mathcal{S}},Z_1) \in [2^{\hat{c}_\text{b} n}]$ for all $e \in \outset{2}$, where $Z_1$ is independent of all the source messages $\bm{X}_{\mathcal{S}}$ and has the same distribution as $\hat{Z}$.
    \item Set $g_{t_i} = \hat{g}_i$ for all $i \in \hat{\mathcal{T}}$, and $g_u=0$ for all other vertices.
    \end{itemize}

    For each vertex $t_i$, $i \in \hat{\mathcal{T}}$, all incoming edges $\inset{t_i}$ originate from vertex~2 and vertices $\{s_j: j \in \hat{\mathcal{H}}_i\}$. The message on edge $(2, t_i)$ is $\hat{f}(\bm{X}_{\mathcal{S}},Z_1)$, and that on edge $(s_j , t_i)$ is $f_{(s_j , t_i)} = X_{s_j}$.
    This means vertex $t_i$, $i \in \hat{\mathcal{T}}$, can decode
    \begin{subequations}
      \begin{align} g_{t_i}(\hat{f}(\bm{X}_{\mathcal{S}},Z_1),\bm{X}_{\hat{\mathcal{H}}_{i}}) &=\hat{g}_i(\hat{f}(\bm{X}_{\mathcal{S}},Z_1),\bm{X}_{\hat{\mathcal{H}}_{i}})\\                                                       &= \bm{X}_{\hat{\mathcal{W}}_i} = \bm{X}_{\{j \in \mathcal{S}: t_i \in \mathcal{D}(j)\}}. \label{eq:g}
    \end{align}
  \end{subequations}
  Here, \eqref{eq:g} follows from \eqref{eq:decoding} as $\hat{Z}$ and $Z_1$ have the same distribution and both are independent of the respective sources messages.
    Noting that the rest of the vertices are not the destination for any source message, the network code satisfies the decoding requirements of $I$.

    Each eavesdropper $r \in \mathcal{R} = \hat{\mathcal{R}}$ has access to messages on the edge set $\mathcal{B}_r$ consisting of
    \begin{itemize}
    \item edge $(1, 2)$, which carries $\hat{f}(\bm{X}_{\mathcal{S}},Z)$, and
      \item edges $\{ \outset{s_i}: i \in \hat{\mathcal{B}}_r\}$, which carry messages $\bm{X}_{\hat{\mathcal{B}}_r}$.
      \end{itemize}
      Now,
      \begin{subequations}
      \begin{align}
                H(\bm{X}_{\mathcal{A}_r}| \bm{X}_{\mathcal{B}_r}) &= H(\bm{X}_{\hat{\mathcal{A}}_r}| \hat{f}(\bm{X}_{\mathcal{S}},Z),  \bm{X}_{\hat{\mathcal{B}}_r}) \\
&= H(\bm{X}_{\hat{\mathcal{A}}_{r}}) = H(\bm{X}_{\mathcal{A}_r}), \label{eq:follow}
      \end{align}
    \end{subequations}
    where \eqref{eq:follow} follows from \eqref{eq:security} with a change of variables (from non-hatted to hatted).

      This completes the proof that the network code is also secure against all the eavesdroppers described by $W$.

    \noindent \underline{$I$ is $(\bm{R}_{\mathcal{S}},n)$-feasible  $\Rightarrow$  $\hat{I}$ is $(\hat{\bm{R}}_{\hat{\mathcal{S}}},\hat{c}_{\text{b}},n)$-feasible:}

    Let $\{\mathcal{F}, \mathcal{G}\}$ be a secure network code that supports $(\bm{R}_{\mathcal{S}},n)$. Then,
    \begin{align}
       H(\bm{X}_{\{s \in \mathcal{S}: u \in \mathcal{D}(s)\}}| \bm{X}_{\inset{u}}, X_{O^{-1}(u)}) &=0, \quad u \in \mathcal{V}, \label{eq:decodability-2}
       \end{align}
       and
       \begin{align}
          H(\bm{X}_{\mathcal{A}_r}| \bm{X}_{\mathcal{B}_r}) &= H(\bm{X}_{\mathcal{A}_r}|\{f_e:e \in \mathcal{B}_r\}) \notag\\ &= H(\bm{X}_{\mathcal{A}_r}), \quad r \in \mathcal{R}. \label{eq:decodability-3}
    \end{align}
  By  the construction of $I$, we know that $\{s \in \mathcal{S}: u \in \mathcal{D}(s)\} = \emptyset$, for any $u \in \{s_1, s_2, \dotsc, s_{k},1,2 \}$, and $\{s \in \mathcal{S}: t_i \in \mathcal{D}(s)\} = \hat{\mathcal{W}}_i$ for $i \in \{1,2,\dotsc, \ell\}$. Also, none of the vertices in $\{t_1, t_2, \dotsc, t_{\ell}\}$ is the originating node for any source message. So, \eqref{eq:decodability-2} becomes
    \begin{equation}
      H(\bm{X}_{\hat{\mathcal{W}}_i}| \bm{X}_{\inset{t_i}}) = H(\bm{X}_{\hat{\mathcal{W}}_i}| X_{(2 , t_i)}, X_{l_1}, \dotsc, X_{l_L}) =  0, \label{eq:nc-decoding-function-1}
    \end{equation}
    for $i \in [\ell]$, where $\inset{t_i} = \{(2 , t_i),l_1, l_2, \dotsc, l_L\}$, $L = |\hat{\mathcal{H}}_i|$ and
    \begin{align}
      X_{(2, t_i)} &= f_{2 \rightarrow t_i}(\bar{f}_{(2, t_i)}(\bm{X}_{\mathcal{S}},\bm{Z}), Z_2) \\
      \bm{X}_{\{l_1,\dotsc,l_L\}} &= [f_{(s_{h_1} , t_i)} (X_{h_1},Z_{s_{h_1}}),\notag \\
      &\qquad\qquad \dotsc, f_{(s_{h_L}, t_i)} (X_{h_L},Z_{s_{h_L}})], 
    \end{align}
    where $\hat{\mathcal{H}}_i = \{h_1, \dotsc, h_L\}$, $\bm{Z}$ is the collection of all $\{Z_i\}$ generated by nodes $\{1,s_1, \dotsc, s_{|\mathcal{S}|}\}$, and $Z_2$ is independent of all messages and $\bm{Z}$.

    Decoding at all $t_i$'s must succeed % that \eqref{eq:nc-decoding-function-1} is the average of the conditional entropy over all $\{Z_i\}$ therein, and the conditional entropy must be non-negative
    for any realisation of $\{Z_j\}$. Hence, \eqref{eq:nc-decoding-function-1} must also hold when all $Z_{s_j}=0$. This gives
    \begin{equation}
      H(\bm{X}_{\hat{\mathcal{W}}_i}| X'_{(2 , t_i)}, X'_{l_1}, \dotsc, X'_{l_L}) =  0, \label{eq:nc-decoding-function}
    \end{equation}
    where
    \begin{align}
      X'_{(2 , t_i)} &= f_{(2, t_i)}(\bar{f}_{(1 , 2)}(\bm{X}_{\mathcal{S}},Z_1), Z_2), \\
      \bm{X}'_{\{\ell_1,\dotsc,\ell_L\}} &= [f_{(s_{h_1}, t_i)} (X_{h_1},0), \dotsc, f_{(s_{h_L} , t_i)} (X_{h_L},0)], \label{eq:nc-function2}
    \end{align}
    by setting $Z_{s_i}=0$ for all vertices $s_i$, $i=1,\ldots, k$.

    % This means, there exists some function
    % \begin{align}
    %   g_{t_i}\Big( & f_{2 \rightarrow t_i}(\bar{f}_{1 \rightarrow 2}(\bm{X}_{\mathcal{S}},\bm{Z}), Z_2),f_{\ell_1}(X_{O^{-1}(\tail{\ell_1})},Z_{O^{-1}(\tail{\ell_1})}), \notag \\
    %                &\dotsc, f_{\ell_L}(X_{O^{-1}(\tail{\ell_L})},Z_{O^{-1}(\tail{\ell_L})})\Big) \notag\\
    %   &= \bm{X}_{\mathcal{W}_i},  \label{eq:function}
    % \end{align}
    % where $\bm{Z}$ is the collection of all $\{Z_i\}$ in the
    % Note that \eqref{eq:function} must hold for any realisation of $\{Z_i\}$, which are independent of the source messages.

    Now, we construct a secure index code as follows:
    \begin{equation}
      \hat{X}_{\text{b}} = \hat{f}(\hat{\bm{X}}_{\hat{\mathcal{S}}},\hat{Z}) = \bar{f}_{(1, 2)}(\hat{\bm{X}}_{\hat{\mathcal{S}}},\hat{Z}),
    \end{equation}
    where $\hat{Z} \in [2^{\hat{c}_{\text{b}}n}]$ is a random variable independent of the messages, and having the same distribution as $Z_1$.

    % Also, note that, by construction,
  %   \item $\hat{g}_i(\hat{X}_{\hat{c}_{\text{b}}},\hat{\bm{X}}_{\hat{\mathcal{H}_i}}) = g_{t_i}(\hat{X}_{\hat{c}_{\text{b}}},f_{\ell_1}(\hat{X}_{O^{-1}(\tail{\ell_1})},0),\dotsc,$ $f_{\ell_L}(\hat{X}_{O^{-1}(\tail{\ell_L})},0))$ $\stackrel{(a)}{=} \hat{\bm{X}}_{\hat{\mathcal{W}}_i}$, for each $i \in \hat{\mathcal{T}}$, where $\hat{\mathcal{H}}_i = \{O^{-1}(\tail{\ell_1}), \dotsc, O^{-1}(\tail{\ell_L})\}$, where $\inset{t_i} = \{2, \ell_1, \ell_2, \dotsc, \ell_L\}$. Equality $(a)$ follows by noting that \eqref{eq:function} must hold for any realisations of $\{Z_{\ell_j}\}$.
  %   \end{itemize}
    Now, for each receiver $\hat{t} \in \hat{\mathcal{T}}$,
    \begin{subequations}
      \begin{align}
        &H(\hat{\bm{X}}_{\hat{\mathcal{W}}_{\hat{t}}} | \hat{X}_{\text{b}}, \hat{\bm{X}}_{\hat{\mathcal{H}}_{\hat{t}}}) \notag \\
        &= H(\hat{\bm{X}}_{\hat{\mathcal{W}}_{\hat{t}}} | \hat{X}_{\text{b}}, \hat{\bm{X}}_{\hat{\mathcal{H}}_{\hat{t}}}, f_{(s_{h_1}, t_{\hat{t}})} (\hat{X}_{h_1},0), \dotsc, f_{(s_{h_L}, t_{\hat{t}}}) (\hat{X}_{h_L},0) ) \notag \\
        &= H(\hat{\bm{X}}_{\hat{\mathcal{W}}_{\hat{t}}} | \hat{X}_{\text{b}}, \hat{\bm{X}}_{\hat{\mathcal{H}}_{\hat{t}}}, f_{(s_{h_1} , t_{\hat{t}})} (\hat{X}_{h_1},0), \dotsc, f_{(s_{h_L} , t_{\hat{t}})} (\hat{X}_{h_L},0), \notag \\
        &\quad \quad f_{(2 , t_{\hat{t}})}(\hat{X}_{\text{b}},\hat{Z}_2) ) \label{eq:markov-nc} \\
        & \leq H(\hat{\bm{X}}_{\hat{\mathcal{W}}_{\hat{t}}} |  f_{(s_{h_1} , t_{\hat{t}})} (\hat{X}_{h_1},0),  \dotsc, f_{(s_{h_L}, t_{\hat{t}})} (\hat{X}_{h_L},0), \nonumber\\
&\quad\quad
        f_{(2 , t_{\hat{t}})}(\hat{X}_{\text{b}},\hat{Z}_2) ) \notag \\
        &= 0, \label{eq:nc-decoding-done}
      \end{align}
    \end{subequations}
    where \eqref{eq:nc-decoding-done} follows from \eqref{eq:nc-decoding-function} with a change of variables (from hatted to non-hatted), and
    \eqref{eq:markov-nc} follows from the following Markov chains
    \begin{multline}
      \hat{\bm{X}}_{\hat{\mathcal{W}}_{\hat{t}}} - ( \hat{X}_{\text{b}}, \hat{\bm{X}}_{\hat{\mathcal{H}}_{\hat{t}}}, f_{s_{h_1} \rightarrow t_{\hat{t}}} (\hat{X}_{h_1},0), \dotsc, f_{s_{h_L} \rightarrow t_{\hat{t}}} (\hat{X}_{h_L},0))\\ - f_{2 \rightarrow t_{\hat{t}}}(\hat{X}_{\text{b}},\hat{Z}_2),
    \end{multline}
    as $\hat{Z}_2$ is independent of all other random variables, and has the same distribution as $Z_2$.

  Since conditional entropy is non-negative, it follows from \eqref{eq:nc-decoding-done} that  each receiver $\hat{t} \in \hat{\mathcal{T}}$ can decode the messages $\hat{\bm{X}}_{\hat{\mathcal{W}}_{\hat{t}}}$ that it requires, given $(\hat{X}_{\text{b}}, \hat{\bm{X}}_{\hat{\mathcal{H}}_{\hat{t}}})$.

  We now prove the security constraints in $\hat{I}$. First, consider $I$. If $X_i$ is requested by some nodes, then observing all outgoing links from $s_i$ must enable one to reconstruct $X_i$. If $X_j$ is not requested by any node,  we assume that observing all outgoing links from $s_j$ also enables one to reconstruct $X_j$. The rationale behind this assumption is as follows: If $R_j=R'$ is in the feasible region, then all $R_j > R'$ are also in the feasible region by having node $s_j$ transmitting the first $nR'$ bits of $X_j$. Hence, we only need to consider the smallest feasible rate for $X_j$, denoted by $R_\text{min}$, when all the other rates are kept fixed. Now, if after observing all outgoing links from $s_j$, one can obtain only $nR_j''$ bits of information of $X_j$ (where $R_j'' < R_\text{min}$), then node $s_j$ could have transmitted $X_j$ at rate $R_j''$, which contradicts that $R_\text{min}$ is the smallest feasible rate for $X_j$.

  So, from \eqref{eq:decodability-3}, we have
  \begin{subequations}
    \begin{align}
      \hspace{-2mm}H(\bm{X}_{\mathcal{A}_r}) &= H(\bm{X}_{\mathcal{A}_r} | \{ X_e: e \in \mathcal{B}_r\}) \label{eq:independent-2} \\
      &= H(\bm{X}_{\mathcal{A}_r} | \bar{f}_{(1, 2)}(\bm{X}_{\mathcal{S}},Z_{1}), \bm{X}_{\{\outset{s_i} : i \in \hat{\mathcal{B}}_r\}}) \label{eq:independent-3}\\
      &= H(\bm{X}_{\mathcal{A}_r} | \bar{f}_{(1, 2)}(\bm{X}_{\mathcal{S}},Z_{1}), \bm{X}_{\{\outset{s_i} : i \in \hat{\mathcal{B}}_r\}}, \bm{X}_{\hat{\mathcal{B}}_r} ) \label{eq:function-001}\\
&\leq H(\bm{X}_{\mathcal{A}_r} | \bar{f}_{(1, 2)}(\bm{X}_{\mathcal{S}},Z_{1}), \bm{X}_{\hat{\mathcal{B}}_r} )\label{eq:function-002}\\
&\leq H(\bm{X}_{\mathcal{A}_r}),\label{eq:function-003}
    \end{align}
  \end{subequations}
where \eqref{eq:function-001} is derived because observing all outgoing links from $s_i$ allows one to reconstruct $X_i$. Thus, it follows that \eqref{eq:function-002} and \eqref{eq:function-003} must hold with equality.

  Now, consider an eavesdropper $\hat{r} \in \hat{\mathcal{R}}$ in the index-coding equivalence $\hat{I}$.
  \begin{subequations}
  \begin{align}
    H(\hat{\bm{X}}_{\hat{\mathcal{A}}_{\hat{r}}} | \hat{\bm{X}}_{\text{b}}, \hat{\bm{X}}_{\hat{\mathcal{B}}_{\hat{r}}})& = H( \hat{\bm{X}}_{\hat{\mathcal{A}}_{\hat{r}}} |  \bar{f}_{(1, 2)}(\hat{\bm{X}}_{\hat{\mathcal{S}}},\hat{Z}), \hat{\bm{X}}_{\hat{\mathcal{B}}_{\hat{r}}}) \notag \\
    &= H(\hat{\bm{X}}_{\mathcal{A}_{\hat{r}}}) = H(\hat{\bm{X}}_{\hat{\mathcal{A}}_{\hat{r}}}), \label{eq:ic-independent}
  \end{align}
\end{subequations}
where
\eqref{eq:ic-independent} follows from \eqref{eq:independent-2}--\eqref{eq:function-003} with a change of variables (from non-hatted to hatted), by noting that $(\hat{\bm{X}}_{\hat{\mathcal{S}}},\hat{Z})$ and $(\bm{X}_{\mathcal{S}},Z_{1})$ have the same distribution. %Note that for the replacement $\hat{X}_i \rightarrow \hat{u}_i$ if $\mathcal{D}(i)=\emptyset$, $i$ does not appear in any $\hat{W}_{\hat{t}}$. Hence, these replacements will not affect the decodability condition~\eqref{eq:nc-decoding-done}.

So, $\hat{X}_\text{b}$ is a secure index code for $\hat{I}$.
\end{IEEEproof}

\section{Mapping from Secure Network Coding to\\
 Secure Index Coding}

Given a secure network-coding instance $\mathcal{I} = ( G, M, W)$, we first construct an augmented secure network-coding instance with deterministic encoding, and then construct an equivalent  secure index-coding instance $\hat{I} = (\hat {\mathcal{S}}, \hat{\mathcal{T}}, \{\hat{\mathcal{W}}_{\hat{t}}\}, \{ \hat{\mathcal{H}}_{\hat{t}} \}, \hat{W} )$.

\noindent \underline{Augmented secure network coding}:
We construct an \emph{augmented} secure network-coding instance $I' = ( G', M', W')$ as follows:
\begin{itemize}
\item $G'= (\mathcal{V}',\mathcal{E}')=G=(\mathcal{V},\mathcal{E})$, and $c'_e = c_e$ for all $e \in \mathcal{E}'$. The vertices, the edges, and the edge capacities remain the same.
\item Let $\mathcal{S} = [S]$, where $S \triangleq |\mathcal{S}|$. The
  connection requirement is augmented as follows: $\mathcal{S}' = \mathcal{S} \cup \{S+1, S+2, \dotsc, S+|\mathcal{V}'|\}$, where we introduce an additional source $X'_{S+v} \in [\prod_{e \in \outset{v}} 2^{c_en}] \triangleq [2^{k_v n}]$, originating at each vertex $v \in [|\mathcal{V}'|]$, that takes the role of and has the same distribution as the random key $Z_v$ used in the randomised encoding at vertex~$v$ in $I$. So, $O'(S+v) = v$, i.e., $X'_{S+v}$ originates at vertex $v$, and we define $\mathcal{D}'(S+v) = \emptyset$, i.e., $X'_{S+v}$ is not requested by any vertex.
 For $s \in \mathcal{S}$, $\bm{X}'_{\mathcal{S}}$ has the same distribution as $\bm{X}_{\mathcal{S}}$, $O'(s) = O(s)$, and $\mathcal{D}'(s) = \mathcal{D}(s)$.  Note that $R'_s = R_s$ for all $s \in [S]$, and $R'_{S+v} = k_v$ for all $v \in [|\mathcal{V}'|]$.
  \item $W'=W$, i.e., $\mathcal{R}' = \mathcal{R}$, $\mathcal{B}'_r = \mathcal{B}_r$, and $\mathcal{A}'_r = \mathcal{A}_r$. The adversarial setting remains the same. Thus, messages $\{X'_{S+v}: v \in [|\mathcal{V}']\}$ are neither known to the adversaries nor need to be protected.
  %\item Note that the messages $\{X_s: s \in \{S+1, \dotsc, S+|\mathcal{V}'|\}\}$ are mutually independent and are independent of all other messages, but they need not be uniformly distributed.
  \end{itemize}
  Any deterministic or randomised (i.e., using an independent random key
  $Z_v$ at vertex $v$) secure network code for $I$ is equivalent to a
  deterministic secure network code for $I'$, where each node $v$ gets an
  additional source $X'_{S+v}$ that is not required to be decoded by any node.

%  Since $\{Z_v\}$ and $\{X_{S+v}\}$ play the role of random keys for security in secure network coding, we assume that they are each uniformly distributed over an alphbet of size $2^{k_v n}$ respecively, where $2^{k_v n} \leq \prod_{e \in \outset{v}} 2^{c_en}$. It follows that $R'_s = R_s$ for $s \in [S]$ and $R'_{S+v} = k_v$ for $v \in |\mathcal{V}'|]$.

  Denote the set of vertices in $\mathcal{I}'$ that are destinations for some source messages by $\mathcal{T}' = \{j \in \mathcal{V}': j \in \mathcal{D}'(i) \text{ for some } i \in \mathcal{S}'\}$. Note that $O'(\cdot)$ can map different source indices to one vertex, and hence, $O'^{-1}(j)$ returns a set of indices of messages originating at vertex~$j$.

\begin{figure*}
  \centering
  \begin{subfigure}[b]{0.17\textwidth}
    \centering
    \resizebox{23ex}{!}{%
\includegraphics{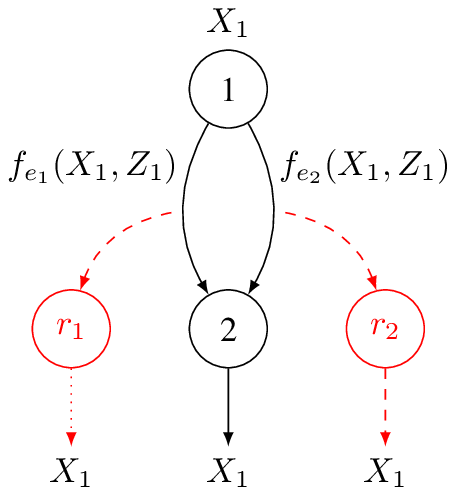}
% \begin{tikzpicture} [node distance=5ex,
% receiver/.style={circle,draw, minimum size=5ex},
% point/.style={circle,inner sep=0pt}
% ]
% \node (1) [receiver,label=above:{$X_1$}] {1};
% \node (m) [point, below=of 1] {};
% \node (2) [receiver,below=of m] {2};
% \draw[->,>=latex, bend left] (1) to node[near start,right] (es2) {$f_{e_2}(X_1,Z_1)$} node[midway] (em2) {}  (2);
% \draw[->,>=latex, bend right] (1) to node[near start,left] (es1) {$f_{e_1}(X_1,Z_1)$} node[midway] (em1) {}  (2);

% \node (v1) [receiver, left=of 2, color=red] {\color{red}$r_1$};
% \node (v2) [receiver, right=of 2,color=red] {\color{red}$r_2$};

% \node (bv1) [point, below=of v1] {$X_1$};
% \node (b2) [point, below=of 2] {$X_1$};
% \node (bv2) [point, below=of v2] {$X_1$};

% \draw[->,>=latex, color=red, dotted] (v1) to (bv1);
% \draw[->,>=latex] (2) to (b2);
% \draw[->,>=latex, color=red, dashed] (v2) to (bv2);
% \draw[->,>=latex, color=red, dashed, bend right] (em1) to (v1);
% \draw[->,>=latex, color=red, dashed, bend left] (em2) to (v2);
% \end{tikzpicture}
}%
        \caption{$I$ with randomised encoding}
        \label{fig:network-coding-1}
      \end{subfigure}
      \quad \quad
  \begin{subfigure}[b]{0.17\textwidth}
    \centering
    \resizebox{23ex}{!}{%
\includegraphics{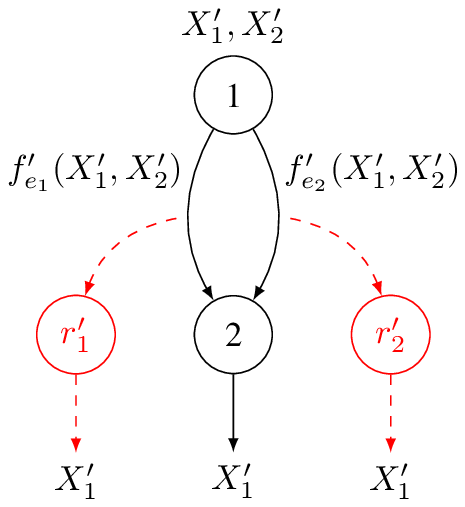}
% \begin{tikzpicture} [node distance=5ex,
% receiver/.style={circle,draw, minimum size=5ex},
% point/.style={circle,inner sep=0pt}
% ]
% \node (1) [receiver,label=above:{$X'_1,X'_2$}] {1};
% \node (m) [point, below=of 1] {};
% \node (2) [receiver,below=of m] {2};
% \draw[->,>=latex, bend left] (1) to node[near start,right] (es2) {$f'_{e_2}(X'_1,X'_2)$} node[midway] (em2) {}  (2);
% \draw[->,>=latex, bend right] (1) to node[near start,left] (es1) {$f'_{e_1}(X'_1,X'_2)$} node[midway] (em1) {}  (2);

% \node (v1) [receiver, left=of 2, color=red] {\color{red}$r'_1$};
% \node (v2) [receiver, right=of 2,color=red] {\color{red}$r'_2$};

% \node (bv1) [point, below=of v1] {$X'_1$};
% \node (b2) [point, below=of 2] {$X'_1$};
% \node (bv2) [point, below=of v2] {$X'_1$};

% \draw[->,>=latex, color=red, dashed] (v1) to (bv1);
% \draw[->,>=latex] (2) to (b2);
% \draw[->,>=latex, color=red, dashed] (v2) to (bv2);
% \draw[->,>=latex, color=red, dashed, bend right] (em1) to (v1);
% \draw[->,>=latex, color=red, dashed, bend left] (em2) to (v2);
% \end{tikzpicture}
}
        \caption{$I'$ with deterministic encoding}
        \label{fig:network-coding-2}
      \end{subfigure}
      \quad
    \begin{subfigure}[b]{0.55\textwidth}
\centering
\resizebox{67ex}{!}{%
\includegraphics{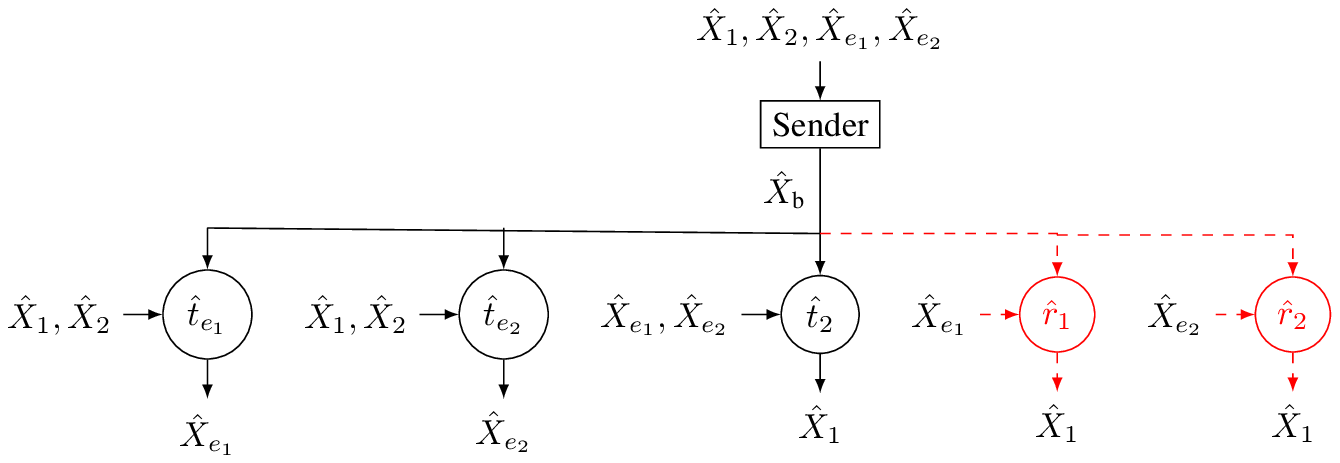}
}%
        \caption{$\hat{I}$ with deterministic encoding}
        \label{fig:index-coding-1}
      \end{subfigure}
      \quad
 
      \caption{A secure network-coding instance ${I}$, its augmented
        version $I'$, and the corresponding secure index-coding instance
        $\hat{I}$, where $r_1, r_2, r'_1, r'_2, \hat{r}_1, \hat{r}_2$ are
        eavesdroppers}
      \vspace{-2ex}
      \label{fig:network-to-index-example}
\end{figure*}

\noindent \underline{Network-to-index coding mapping}:
\begin{itemize}
\item $\hat{\mathcal{S}} = \mathcal{S}' \cup \mathcal{E}'$. It consists of one source message $\hat{X}_s$ for each $s \in \mathcal{S}'$ in $I'$, and one $\hat{X}_e$ for each edge $e \in \mathcal{E}'$ in $I'$. $\hat{\bm{X}}_{\mathcal{S}'}$ has the same distribution as $\bm{X}'_{\mathcal{S}'}$. The rates of the messages are $\hat{R}_s = R'_s$ and $\hat{R}_e = c'_e$.
\item $\hat{\mathcal{T}} = \{\hat{t}_i\}_{i \in \mathcal{T}'} \cup \{\hat{t}_e\}_{e \in \mathcal{E}'}$. This means $\hat{I}$ has $|\mathcal{T}'|+|\mathcal{E}'|$ receivers, one for each destination node in $I'$ and one for each edge in $I'$.
\item For each $\hat{t}_e \in \hat{\mathcal{T}}$, $\hat{\mathcal{H}}_{\hat{t}_e} = \inset{\tail{e}} \cup O'^{-1}(\tail{e})$, and $\hat{\mathcal{W}}_{\hat{t}_e} = \{ e \}$.
\item For each $\hat{t}_i \in \hat{\mathcal{T}}$, $\hat{\mathcal{H}}_{\hat{t}_i} = \inset{i} \cup  O'^{-1}(i)$, and $\hat{\mathcal{W}}_{\hat{t}_i} = \{s \in [S]: i \in \mathcal{D}'(s) \}$.
\item The eavesdropper setting $W'$: $\hat{\mathcal{R}} = \mathcal{R}'$. For each $\hat{r} \in \hat{\mathcal{R}}$, $\hat{\mathcal{B}}_{\hat{r}} = \mathcal{B}'_{\hat{r}}$, and $\hat{\mathcal{A}}_{\hat{r}} = \mathcal{A}'_{\hat{r}}$.
  \item We set the broadcast rate as $\hat{c}_{\text{b}} = \sum_{e \in \mathcal{E}'}c'_e$.
  \end{itemize}

Figure~\ref{fig:network-to-index-example} depicts an example of such a mapping.

  \begin{remark}
    This network-to-index coding mapping is slightly different from that of Effros et al.~\cite{effrosrouayheblangberg15}, since we do not require the use of an additional receiver $\hat{t}_\text{all}$ for the corresponding index-coding instance. We will show that omitting this receiver will not affect the equivalence.
\end{remark}

  Note that unlike the index-to-network mapping, here $\hat{\bm{X}}_{\mathcal{E}'}$ and $\bm{X}'_{\mathcal{E}'}$ have different distributions, where the latter are functions of $\bm{X}'_{\mathcal{S}'}$. For the corresponding secure index-coding instance, we choose $\{\hat{X}_e: e \in \mathcal{E}'\}$ to be mutually independent, independent of all other messages, and each $\hat{X}_e$ is uniformly distributed over $[2^{\hat{R}_e n}]$. We will see that using uniformly distributed $\hat{X}_e$ is the key to ensuring security.

  With the above conversion, we now state an equivalence between $I$ and $\hat{I}$ through $I'$:
  \begin{theorem}\label{theorem:network-to-index}
    Let $I$ be a secure network-coding instance and $I'$ be its augmented instance. Let $\hat{I}$ and $\hat{c}_{\text{b}}$ be the corresponding secure index-coding instance and a broadcast rate, respectively, obtained using the network-to-index coding mapping from $I'$. For any ${\bm{R}}_{\mathcal{S}}$, the instance $I$ is $(\bm{R}_{\mathcal{S}},n)$-feasible if and only if  the instance $\hat{I}$ is $(\hat{\bm{R}}_{\hat{\mathcal{S}}},\hat{c}_{\text{b}},n)$-feasible.
  \end{theorem}

  \begin{IEEEproof}[Proof of Theorem~\ref{theorem:network-to-index}]

   \noindent \underline{$I$ is $(\bm{R}_{\mathcal{S}},n)$-feasible  $\Rightarrow$  $\hat{I}$ is $(\hat{\bm{R}}_{\hat{\mathcal{S}}},\hat{c}_{\text{b}},n)$-feasible:}

   Note that $I$ is $(\bm{R}_{\mathcal{S}},n)$-feasible if and only if $I'$ is $(\bm{R}'_{\mathcal{S}'},n)$-feasible using \textit{deterministic} network encoding functions $\{f'_e\}$ derived from $\{f_e\}$ for $I$, where all the randomness $\{Z_v\}$ in the network code for $I$ is realised using $\{X_{S+v}\}$ in $I'$.

   Since the network code for $I'$ is deterministic, we use the same code mapping as that proposed by Effros et al.~\cite{effrosrouayheblangberg15}: The sender's broadcast message is $\hat{X}_{\text{b}} = [\hat{X}_{\text{b}}(e)]_{e \in \mathcal{E}'} $, where
   \begin{equation}
     \hat{X}_{\text{b}}(e) = \hat{X}_e + \bar{f}'_e(\hat{\bm{X}}_{\mathcal{S}'}). \label{eq:edge-code}
   \end{equation}
   Note that $\hat{X}_e, \bar{f}'_e \in [2^{\hat{R}_e n}] = [2^{c'_e n}] = [2^{c_e n}]$.

   In $I'$, each vertex~$v \in \mathcal{T}'$ can decode all messages that it requires from the message on all incoming edges and messages originating at $v$, meaning that
   \begin{align*}
     \bm{X}'_{\{s \in \mathcal{S}': v \in \mathcal{D}'(s)\}} &= \bm{X}'_{\{s \in [S]: v \in \mathcal{D}'(s)\}} = g'_v(\bm{X}'_{\inset{v} \cup O'^{-1}(v)}) \\ &= g'_v(\bm{X}'_{\inset{v}}, \bm{X}'_{O'^{-1}(v)})\\& = g'_v( [\bar{f}'_e(\bm{X}'_{\mathcal{S}'})]_{e \in \inset{v})}, \bm{X}'_{O'^{-1}(v)}).
   \end{align*}
   
 As mentioned above, while messages $\bm{X}'_{O'^{-1}(v)}$ and $\hat{\bm{X}}_{O'^{-1}(v)}$ (with node subscripts) have the same distribution, messages $\bm{X}'_{\inset{v}}$ and $\hat{\bm{X}}_{\inset{v}}$ (with edge subscripts) may not. 
To deal with this issue, consider the broadcast message $\hat{X}_{\text{b}}$. From \eqref{eq:edge-code}, any receiver that knows $\hat{X}_e$ can obtain $\bar{f}'_e(\hat{\bm{X}}_{\mathcal{S}'})$, where $[\bar{f}'_e(\hat{\bm{X}}_{\mathcal{S}'})]_{e \in \mathcal{E}'}$ and $[\bar{f}'_e(\bm{X}'_{\mathcal{S}'})]_{e \in \mathcal{E}'}$ have the same distribution.

 So, with a change of variables (from non-hatted to hatted), receiver $\hat{t}_i \in \hat{\mathcal{T}}$ can decode the messages it requires using
   \begin{equation*}
     \hat{\bm{X}}_{\hat{\mathcal{W}}_{\hat{t}_i}} = \hat{\bm{X}}_{\{s \in [S]: i \in \mathcal{D}'(s) \}} = g'_i( [\bar{f}'_e(\hat{\bm{X}}_{\mathcal{S}'})]_{ e \in \inset{i})}, \hat{\bm{X}}_{O'^{-1}(i)}).
   \end{equation*}
   As receiver $\hat{t}_i$ knows $\hat{\mathcal{H}}_{\hat{t}_i} = \inset{i} \cup  O'^{-1}(i)$ by the mapping, it knows $\hat{\bm{X}}_{O'^{-1}(i)}$ and can obtain $[\bar{f}'_e(\hat{\bm{X}}_{\mathcal{S}'})]_{ e \in \inset{i})}$ from $\hat{X}_\text{b}$ and $\hat{\bm{X}}_{\inset{i}}$. % This is possible if $\hat{\bm{X}}_{\mathcal{S}'}$ and $\hat{\bm{X}}_{\mathcal{S}'}$ have the same distribution.

Receiver $\hat{t}_e \in \hat{\mathcal{T}}$ uses \eqref{eq:edge-code} to obtain the required $\hat{X}_e$ from $\hat{X}_{\text{b}}(e) - \bar{f}'_e(\hat{\bm{X}}_{\mathcal{S}'})$, where the first term is the broadcast message available to the receiver $\hat{t}_e$. To obtain the second term, express the global encoding function as its local encoding function, $\bar{f}'_e(\hat{\bm{X}}_{\mathcal{S}'}) = f'_e([\bar{f}'_{e'}(\hat{\bm{X}}_{\mathcal{S}'})]_{e' \in \inset{\tail{e}}},\hat{\bm{X}}_{O'^{-1}(\tail{e})})$, where $\hat{\bm{X}}_{O'^{-1}(\tail{e})}$ is available to receiver $\hat{t}_e$ as side information. From the broadcast message, receiver $\hat{t}_e$ can obtain
%   \begin{equation}
     $\bar{f}'_{e'}(\hat{\bm{X}}_{\mathcal{S}'}) = \hat{X}_{\text{b}}(e') - \hat{X}_{e'},$
%     \end{equation}
     as it has $\hat{X}_{e'}$, $e' \in \inset{\tail{e}},$ as side information. With this, we have shown that each $\hat{t} \in \hat{\mathcal{T}}$ can decode the messages that it requires.

     We now consider the security constraints. % Note that, from the mapping, $\hat{\mathcal{B}}_{\hat{r}}$ contains only edge indices. Knowing only edges messages $\{\hat{X}_e: e \in \hat{\mathcal{B}}_{\hat{r}}\}$ and the broadcast message $\hat{X}_{\text{b}}$, eavesdropper~$\hat{r}$ can only obtain $\{\bar{f}'_e(\hat{\bm{X}}_{\mathcal{S}'}):e \in \hat{\mathcal{B}}_{\hat{r}}\}$, as $\{\bar{f}'_{e'}(\hat{\bm{X}}_{\mathcal{S}'}): e' \notin \hat{\mathcal{B}}_{\hat{r}}\}$ have been randomised by independently and uniformly distributed $\hat{X}_{e'}$. So,
     For each $\hat{r} \in \hat{\mathcal{R}}$,
     \begin{subequations}
       \begin{align}
         &H(\hat{\bm{X}}_{\hat{\mathcal{A}}_{\hat{r}}} | \hat{X}_{\text{b}}, \hat{\bm{X}}_{\hat{\mathcal{B}}_{\hat{r}}}) \\
         &= H(\hat{\bm{X}}_{\hat{\mathcal{A}}_{\hat{r}}} | \{\hat{X}_{\text{b}}(e):e \in \mathcal{E}'\}, \{ \hat{X}_{e'}: e' \in \hat{\mathcal{B}}_{\hat{r}} \}) \\
         &= H(\hat{\bm{X}}_{\hat{\mathcal{A}}_{\hat{r}}} | \{\hat{X}_{\text{b}}(e):e \in \hat{\mathcal{B}}_{\hat{r}}\}, \{ \hat{X}_{e'}: e' \in \hat{\mathcal{B}}_{\hat{r}} \}) \label{eq:blocked} \\
         &= H(\hat{\bm{X}}_{\hat{\mathcal{A}}_{\hat{r}}} | \{\hat{X}_{\text{b}}(e), \hat{X}_{e'}, \bar{f}'_e(\hat{\bm{X}}_{\mathcal{S}'}):e \in \hat{\mathcal{B}}_{\hat{r}}\}) \label{eq:edge-function-2} \\
         & = H(\hat{\bm{X}}_{\hat{\mathcal{A}}_{\hat{r}}} | \{ \bar{f}'_e(\hat{\bm{X}}_{\mathcal{S}'}):e \in \hat{\mathcal{B}}_{\hat{r}}\}) \label{eq:must-be-equality} \\
         &= H(\hat{\bm{X}}_{\mathcal{A}'_{\hat{r}}} | \{ \bar{f}'_e(\hat{\bm{X}}_{\mathcal{S}'}):e \in \mathcal{B}'_{\hat{r}}\})\\
         &= H(\hat{\bm{X}}_{\mathcal{A}'_{\hat{r}}}) = H(\hat{\bm{X}}_{\hat{\mathcal{A}}_{\hat{r}}}), \label{eq:change-var}
       \end{align}
     \end{subequations}
    where \eqref{eq:blocked} follows from the Markov chain
\begin{multline*}
    \hat{\bm{X}}_{\hat{\mathcal{A}}_{\hat{r}}} - \left(\{\hat{X}_{\text{b}}(e):e
    \in \hat{\mathcal{B}}_{\hat{r}}\}, \{ \hat{X}_{e'}: e' \in
    \hat{\mathcal{B}}_{\hat{r}} \}\right)\\ - (\{\hat{X}_{\text{b}}(e):e \notin
    \hat{\mathcal{B}}_{\hat{r}}\}),
  \end{multline*}
  where  $\{\hat{X}_{\text{b}}(e):e \notin \hat{\mathcal{B}}_{\hat{r}}\}$ has been randomised by independently and uniformly distributed $\{\hat{X}_e:e \notin \hat{\mathcal{B}}_{\hat{r}}\}$, which are independent of $(\hat{\bm{X}}_{\hat{\mathcal{A}}_{\hat{r}}}, \hat{\bm{X}}_{\hat{\mathcal{B}}_{\hat{r}}},\hat{\bm{X}}_{\mathcal{S}'})$ (see \eqref{eq:edge-code}); \eqref{eq:edge-function-2} follows from \eqref{eq:edge-code}; \eqref{eq:must-be-equality} follows from the Markov chain
$$\hat{\bm{X}}_{\hat{\mathcal{A}}_{\hat{r}}} - \{ \bar{f}'_e(\hat{\bm{X}}_{\mathcal{S}'}):e \in \hat{\mathcal{B}}_{\hat{r}}\} - \{ \hat{X}_{e'},\hat{X}_{\text{b}}(e):e \in \hat{\mathcal{B}}_{\hat{r}}\},$$
which can be derived from \eqref{eq:edge-code} and noting that $\{\hat{X}_e: e \in \mathcal{E}'\}$ are independent of $(\hat{\bm{X}}_{\hat{\mathcal{A}}_{\hat{r}}}, \hat{\bm{X}}_{\hat{\mathcal{B}}_{\hat{r}}},\hat{\bm{X}}_{\mathcal{S}'})$; and \eqref{eq:change-var} follows from \eqref{eq:secure-nc-condition} by a change of variables (from hatted to non-hatted) and noting that $\{ \bar{f}'_e(\bm{X}'_{\mathcal{S}'}):e \in \mathcal{B}'_{\hat{r}}\} = \bm{X}'_{\mathcal{B}'_{\hat{r}}}$

    So, the index code is secure.

    \noindent \underline{$\hat{I}$ is $(\hat{\bm{R}}_{\hat{\mathcal{S}}},\hat{c}_{\text{b}},n)$-feasible $\Rightarrow$ $I$ is $(\bm{R}_{\mathcal{S}},n)$-feasible:}

    We will show that if $\hat{I}$ is $(\hat{\bm{R}}_{\hat{\mathcal{S}}},\hat{c}_{\text{b}},n)$-feasible, then $I'$ is $(\bm{R}'_{\mathcal{S}'},n)$-feasible, which implies that $I$ is $(\bm{R}_{\mathcal{S}},n)$-feasible.

    Again, we use the network-code construction proposed by Effros et al.~\cite{effrosrouayheblangberg15}. Note that for a secure index code, there exists a decoding function at receiver $\hat{t}_i$ for each $i \in \mathcal{T}'$, such that
    \begin{subequations}
      \begin{align}
        \hat{g}_{\hat{t}_i}(\hat{X}_{\text{b}},\hat{\bm{X}}_{\hat{\mathcal{H}}_{\hat{t}_i}})
        &= \hat{g}_{\hat{t}_i}(\hat{X}_{\text{b}},\hat{\bm{X}}_{\inset{i} \cup  O'^{-1}(i)}) \label{eq:generate-dest}\\
        &= \hat{\bm{X}}_{\hat{\mathcal{W}}_{\hat{t}_i}} = \hat{\bm{X}}_{\{s \in [S]: i \in \mathcal{D}'(s)\}},
      \end{align}
    \end{subequations}
    and a decoding function at receiver $\hat{t}_e$, $e \in \mathcal{E}'$, such that
    \begin{subequations}
      \begin{align}
        \hat{g}_{\hat{t}_e}(\hat{X}_{\text{b}},\hat{\bm{X}}_{\hat{\mathcal{H}}_{\hat{t}_e}})
        &= \hat{g}_{\hat{t}_e}(\hat{X}_{\text{b}},\hat{\bm{X}}_{\inset{\tail{e}} \cup O'^{-1}(\tail{e})})\\
        &= \hat{\bm{X}}_{\hat{\mathcal{W}}_{\hat{t}_e}}= \hat{X}_{e}. \label{eq:generate-edge}
      \end{align}
    \end{subequations}
    In the secure index-coding instance $\hat{I}$, messages $\hat{\bm{X}}_{\mathcal{E}'}$ are independent of messages $\hat{\bm{X}}_{\mathcal{S}'}$, and the broadcast message $\hat{X}_{\text{b}}$ is a function of these messages. However, given $\hat{X}_{\text{b}}$, the messages $\hat{\bm{X}}_{\mathcal{E}'}$ and $\hat{\bm{X}}_{\mathcal{S}'}$ are dependent.

    We set $\hat{X}_{\text{b}}=\sigma$ (which is an arbitrary but valid realisation of $\hat{X}_{\text{b}}$ in the network code) for all $\hat{g}_{\hat{t}_i}(\cdot)$ and $\hat{g}_{\hat{t}_e}(\cdot)$, and choose
    \begin{align}
      X'_e &= g'_e(\bm{X}'_{\inset{\tail{e}} \cup O'^{-1}(\tail{e})}) \\
      &= \hat{g}_{\hat{t}_e}(\sigma,\bm{X}'_{\inset{\tail{e}} \cup O'^{-1}(\tail{e})}), \label{eq:generate-edge-1}
    \end{align}
    for all edges $e \in \mathcal{E}'$, and
    \begin{equation}g'_i(\bm{X}'_{\inset{i} \cup  O'^{-1}(i)}) = \hat{g}_{\hat{t}_i}(\sigma,\bm{X}'_{\inset{i} \cup  O'^{-1}(i)}), \label{eq:generate-dest-1}
    \end{equation}
    for each destination vertex $i \in \mathcal{T}'$. By fixing the first argument in the functions to be $\sigma$, $\bm{X}'_{\mathcal{E}'}$ are now functions of the source messages $\bm{X}'_{\mathcal{S}'}$, and they can be generated following the (acyclic) graph topology of $I'$. Now,  \eqref{eq:generate-dest}--\eqref{eq:generate-edge} hold for any realisation of the variables $\hat{\bm{X}}_{\mathcal{S}' \cup \mathcal{E}'}$. So, for any realisation of the messages $\bm{x}'_{\mathcal{S}'}$, using \eqref{eq:generate-edge-1} with the chosen $\sigma$, and following the topology of $G'$, we can generate the correct and unique realisation of $x'_e$ for every edge~$e$. This will ensure that the decoding step \eqref{eq:generate-dest-1} for each destination $i \in \mathcal{T}'$ gives the correct $\bm{x}'_{\{s \in [S]: i \in \mathcal{D}'(s)\}}$. Thus, correct decoding can be achieved without using the additional receiver $\hat{t}_{\text{all}}$ proposed by Effros et al.~\cite{effrosrouayheblangberg15}.

    Finally, consider the security constraints of $I'$.
    Security for the index code implies that
    \begin{equation*}
      H(\hat{\bm{X}}_{\hat{\mathcal{A}}_{\hat{r}}})
      = H(\hat{\bm{X}}_{\hat{\mathcal{A}}_{\hat{r}}} | \hat{X}_{\text{b}}, \hat{\bm{X}}_{\hat{\mathcal{B}}_{\hat{r}}})
      \leq H(\hat{\bm{X}}_{\hat{\mathcal{A}}_{\hat{r}}} | \hat{X}_{\text{b}})
      \leq H(\hat{\bm{X}}_{\hat{\mathcal{A}}_{\hat{r}}}),
    \end{equation*}
    which implies that $\hat{\bm{X}}_{\hat{\mathcal{A}}_{\hat{r}}}$ and $\hat{\bm{X}}_{\hat{\mathcal{B}}_{\hat{r}}}$ are independent given $\hat{X}_b$. In particular, they are independent given $\hat{X}_b=\sigma$. As $\hat{\bm{X}}_{\mathcal{S}'}$ and $\bm{X}'_{\mathcal{S}'}$ have the same distribution, in the event that $\hat{X}_\text{b} = \sigma$, we see from \eqref{eq:generate-dest}--\eqref{eq:generate-dest-1} that  $p(\bm{x}'_{\mathcal{S}'},\bm{x}'_{\mathcal{E}'}) = p(\hat{\bm{x}}_{\mathcal{S}'},\hat{\bm{x}}_{\mathcal{E}'}|\hat{x}_{\text{b}}=\sigma)$. Since
    \begin{equation}
      I(\hat{\bm{X}}_{\hat{\mathcal{A}}_{r}} ; \hat{\bm{X}}_{\hat{\mathcal{B}}_{r}}| \hat{X}_{\text{b}}=\sigma) =0 =I(\hat{\bm{X}}_{\mathcal{A}_{r}} ; \hat{\bm{X}}_{\mathcal{B}_{r}}| \hat{X}_{\text{b}}=\sigma),
    \end{equation}
    we have $I(\bm{X}'_{\mathcal{A}_r}; \bm{X}'_{\mathcal{B}_r})=0$, which gives the required security constraint \eqref{eq:secure-nc-condition} for $I'$.
  \end{IEEEproof}

%\bibliography{../bib}

\vspace{-0.3ex}

\end{document}